# Impact of phase unwrapping on multi-target acoustic lenses for transcranial holography

D. Attali* [1,2], T. Tiennot* [1,a], M. Tanter[1] and JF. Aubry[1]

***Abstract*—Acoustic lenses have been introduced recently to compensate for the phase distortions induced by the propagation across a human skull for ultrasonic deep-brain stimulation in humans. In this study, we present bifocal lenses that compensate for human skull aberrations and allow simultaneous targeting of multiple structures deep in the brain. We investigated the impact of phase unwrapping in the design of the lenses and how this process improves the distribution of pressure produced in N=5 human skulls for two different spatial arrangements of the targets. The results show that unwrapping the phase computed during the design increases the fidelity of the pressure field generated across the human skulls. The spatial precision is on average improved by 73%, and out-of-target energy deposition is on average reduced by 58%. The results presented in this study highlight the importance of phase unwrapping to optimize the safety and efficacy of future transcranial ultrasound stimulations targeting multiple regions.**

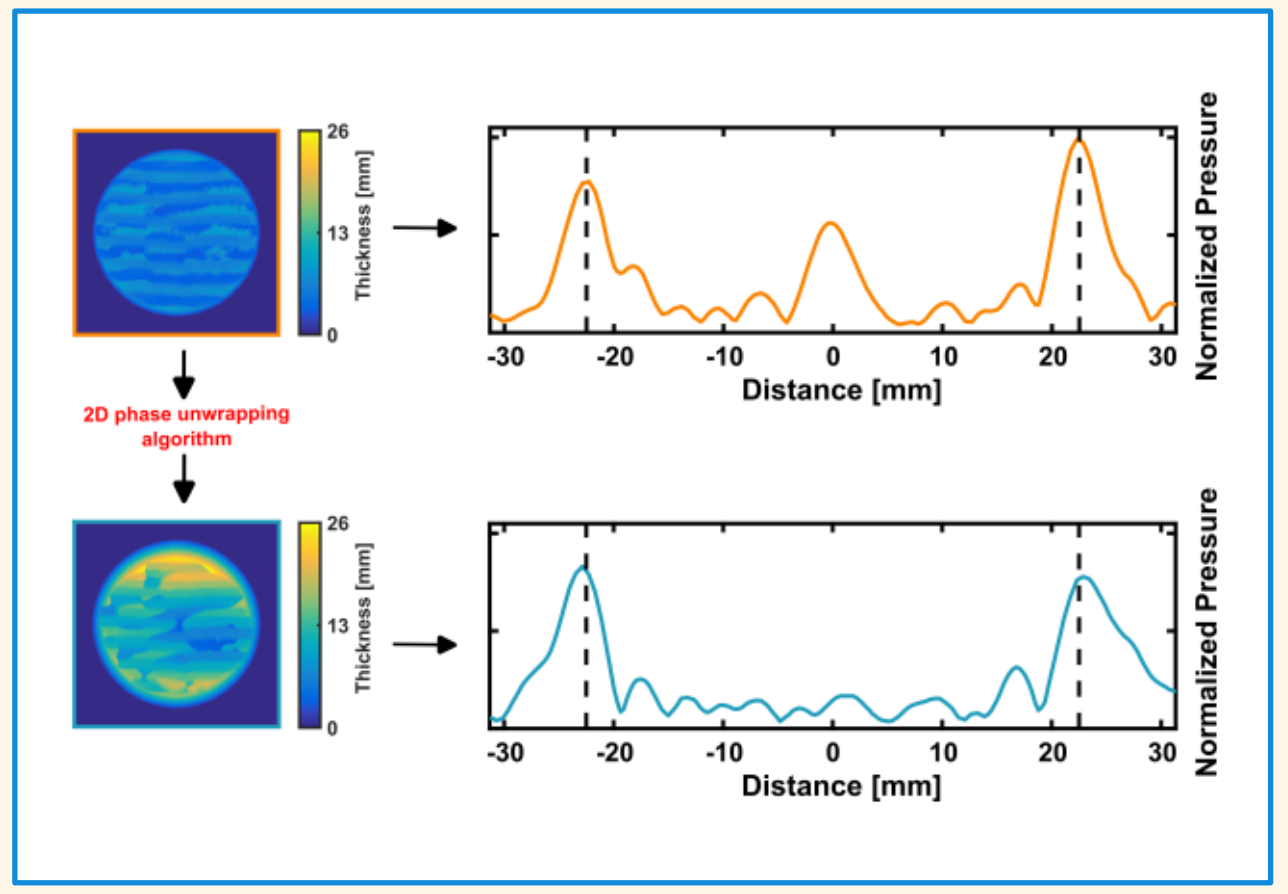

***Index Terms*—Acoustic Lenses; Phase Unwrapping; Transcranial ultrasound stimulation (TUS); Transcranial focused ultrasound (tFUS);**

### *Highlights*

- **Transcranial Ultrasound Stimulation (TUS) is an emerging brain stimulation technique that may benefit from multi-stimulation patterns**
- **We developed acoustic lenses that can both correct for transcranial aberrations and create a bifocal stimulation pattern**
- **We show that phase unwrapping is key in the design of these bifocal acoustic lenses**

## I. Introduction

The ability to create an arbitrary predefined sound field has many applications in different areas of acoustic research like particle trapping [1], [2], cell manipulation for tissue engineering [3] or therapeutical ultrasound [4], [5], [6], [7]. This can be achieved with multielement arrays composed of independently controllable acoustic sources [8], [9], [10]. Recently, a more compact technical solution based on thickness-tuned acoustic lenses and single element transducers has been introduced [8], [11], [12], [13], [14], [15]. Acoustic lenses were first introduced to shape the acoustic field to a predefined target volume for ultrasonic hyperthermia in non-aberrating abdominal tissues [16], [17], [18]. Acoustic lenses were more recently used to account for the phase distortions induced by the propagation across a human skull for ultrasonic transcranial applications [19], [20], [21], [22] and are particularly well suited for low intensity ultrasound applications such as blood-brain barrier opening [23], [24] neuromodulation [25], [26] or both [27], [28]. These applications would benefit from targeting multiple brain structures simultaneously as done in deep brain electrical stimulation protocols [29], [30], [31]. Multi-target therapy strategies are attractive from a neuroscience perspective because they enable modulating brain networks at multiple nodes as opposed to only interacting with a single node using a single target.

[1]Inserm U1273, ESPCI Paris, PSL University, CNRS UMR 8063
Corresponding author: Thomas Tiennot (thomas.tiennot@sonomind.com)
[2]GHU-Paris Psychiatrie et Neurosciences, Hôpital Sainte Anne, Université Paris Cité, 75014 Paris, France

For this reason, several studies have explored using acoustic lenses for multiple targets, first in water [12], [16], [17], [32], [33] and more recently transcranially [21], [22], [24]. However, these proof of concept studies were limited to a single distance between the intended targets, and were performed either through a plastic skull phantom [21] or a single human skull [22], [24], making it difficult to evaluate the robustness of the design of the lens in a larger sample of human skulls.

Here we highlight the limits of simultaneous transcranial targeting with conventional acoustic lenses. We evaluate the performance of conventional lenses on a sample of N=5 human skulls for two different spatial arrangements of the targets. We additionally investigate the impact of phase unwrapping during the design of the lenses and how this process affects the reliability and efficiency of the corresponding acoustic lenses.

## II. Material and Methods

### A. Skull preparation and imaging

Human skulls were provided by the Institute of Anatomy (UFR Biomédicale des Saints-Pères, Université Paris Descartes, Paris, France) and tattooed with individual numbers, as approved by the Ethics Committee of the Centre du Don des Corps (Université Paris Descartes, Paris, France). Prior to each CT scan and acoustic measurement, the skulls were immersed in water and degassed for at least 48 h under a 20 mBar reduced pressure using a desiccator and vacuum pump (LABOPORT N823.3FT.18,KNF, Trenton, New Jersey, USA). CT images were acquired at the GHU Paris Psychiatrie & Neurosciences, Paris, France (Revolution EVO, GE Medical Systems, Chicago, USA). The in-plane spatial resolution of the slices was 0.625 mm. The slices were overlapping, thickness and interslice spacing being 1.25 mm and 0.625 mm, respectively. Voxels with values below 0 Hounsfield units (HU) were set to HUmin = 0, the value of water. Voxels with values above 2400 were set to HUmax = 2400, which is the expected value for cortical bone according to Marsac et al. [34]. The velocity and density in the medium were then derived using a linear approximation [34] and took into account the 3D heterogeneities of the skull bone captured by the Hounsfield units of the CT scan :

$$c(x,y,z) = c_{\text{water}} + (c_{\text{bone}} - c_{\text{water}}) \times \frac{\text{HU}(x,y,z) - \text{HU}_{\text{min}}}{\text{HU}_{\text{max}} - \text{HU}_{\text{min}}}$$

$$\rho(x,y,z) = \rho + (\rho_{\text{bone}} - \rho_{\text{water}}) \times \frac{\text{HU}(x,y,z) - \text{HU}_{\text{min}}}{\text{HU}_{\text{max}} - \text{HU}_{\text{min}}}$$

The speed of sound in water was set to $c_{\text{water}}$ = 1485 m.s$^{-1}$ at 21°C [35] and we chose the following parameters for cortical bone: $c_{\text{bone}}$ = 3100 m s-1 and $\rho_{\text{bone}}$ = 1900 kg.m$^{-3}$ [35].

### B. Ultrasound probe and electronics

The ultrasound transducer used in this study is spherical (radius of curvature 100 mm, aperture diameter 100 mm, f/D ratio 1) and operates at a nominal frequency of 500 kHz (Imasonic SAS, Voray-sur-l'Ognon, France). During the tank experiments, the transducer was powered by a custom-made electronic system consisting of an arbitrary signal generator and an amplifier.

### C. Numerical design of acoustic lenses

The numerical simulations were carried out using the K-Wave toolbox [36]. An isotropic spatial resolution of 10 points per wavelength in water (PPW) was used. The time step was adjusted to give the simulation a Courant-Friedrichs-Lewy number of 0.1. Simulations were performed using the GPU-accelerated k-Wave code on a Titan RTX graphics card (NVIDIA, Santa Clara, CA, USA) and Matlab 2019b (Mathworks, Natick, MA, USA). The 24GB of graphics card memory allowed us to run the full 3D simulation in about 40 minutes using the above parameters. CT scans were linearly upsampled to match the desired resolution of the numerical simulation. Due to the low amplitude pressure associated with neuromodulation applications, nonlinearities were not included in the simulations. Attenuation was not modelled to minimize the simulation time [37].

To design the bifocal lenses, two virtual sources were placed behind digital skull models based on CT scans. The steady-state phase was recorded on a spherical surface positioned at the location where the thickness of the acoustic lens will be calculated, hereafter referred to as the receiving surface. In this study this surface corresponds to the active surface of the transducer. Two options are considered at this stage. Either the phase map is left wrapped between 0 and 2π, or it is subjected to a 2D phase unwrapping process (Fig. 1). Phase unwrapping [38], [39], [40] is used in many applications (synthetic aperture radar (SAR) [41], magnetic resonance imaging (MRI) [42] or fringe pattern analysis [43]) to convert wrapped phase images into continuous phase images. This process fills in possible 2π phase wraps to provide a continuous phase map. In this study we used the 2D phase unwrapping algorithm introduced by Zhao et al [44]. The unwrapping algorithm is based on solving the Transport of Intensity Equation (TIE) [45]. The residual error on the unwrapped phase distribution is iteratively reduced using a discrete cosine transform to solve the Poisson-like equation used in TIE-based approaches. This numerical scheme provides a fast and efficient method for unwrapping noisy phase maps.

For each of the N=5 skulls, two lateral target separations were considered: 15 mm and 30 mm. The two virtual sources were located 5 cm from the outer table of the skull in the axial direction. Their lateral positions are shown in Fig. 2 (15mm case marked with 1 and 30mm case marked with 2). The lenses were then fabricated as described in [19] using the same silicone (Elite double 8, Zhermack, Badina Polesine, Italy) with a speed of sound of 1000 m.s-1 and a density of 1040 kg.m-3. A third configuration with the targets 45 mm apart was also considered for one of the skulls (see Fig. 5) to further illustrate the effect of phase unwrapping as the number of phase wraps increases.

### D. Pressure fields acquisition

Pressure fields were recorded using a needle hydrophone (HNC-0400, Onda Corp., Sunnyvale, CA, USA). The electrical output signal generated by the hydrophone was amplified (AH-2020-DCBSW, Onda Corp., CA, USA) before being recorded on a digital oscilloscope (HS5, TiePie, Sneek, The Netherlands). The hydrophone was mounted on a 3-axis motorised positioning system (ESP300 and ILS Linear Stages, Newport, CA, USA). All the instruments were synchronised using Matlab (Mathworks, Natick, MA, USA). The transmitted signal consisted of a pulse of 60 microseconds, repeated every 20 milliseconds to avoid interference with reflections. The hydrophone signal was acquired over 200 microseconds and the acquisition start was triggered with the emission pulse. The amplitude was measured over 10 cycles (20 microseconds) in the steady-state part of the signal by calculating the amplitude of the Fourier-transform of the 20-microsecond signal at the excitation frequency. For each hydrophone position, the amplitude of the signal was obtained by averaging 10 measurements. The driving voltage was adjusted to produce a peak negative pressure of 1 MPa in water. The voltage-pressure conversion coefficient of the hydrophone-plus-amplifier channel was obtained by equating the peak negative pressure measured at the focal point with the peak negative pressure measured at the same location on an annually calibrated and certified acoustic tank and hydrophone (Acoustic Measurement Tank and Acertara 805 membrane hydrophone, Longmont, CO, USA). The water temperature of the tank was maintained at 21.2◦C±1.0°C using a circulating chiller (Accel 500 LC, Thermo Fisher Scientific Inc., Waltham, MA USA). The average temperature was used to calculate the speed of sound of water in the lens design process using the relationship from [35].

The transducer was housed in a custom 3D printed holder and mounted on a robotic arm (UR3e, Universal Robots A/S, Odense, Denmark). The same piece was also used to hold the lenses in position in front of the transducer. We first scanned the pressure field in water without the skull to determine the position of the acoustic focus. We then used a 3D-printed holder (see Fig. 1.a) designed for each skull to position its outer table 5 cm away from the front end of the transducer (see Fig. 2), as described in [19], [20]. The skull was then held in place with a stereotactic frame (see Fig. 1.b). The skull holder was then removed before starting the measurements. The position of the acoustic focus in the water was used to define the centre of each 2D scan and its depth (see Fig. 2).

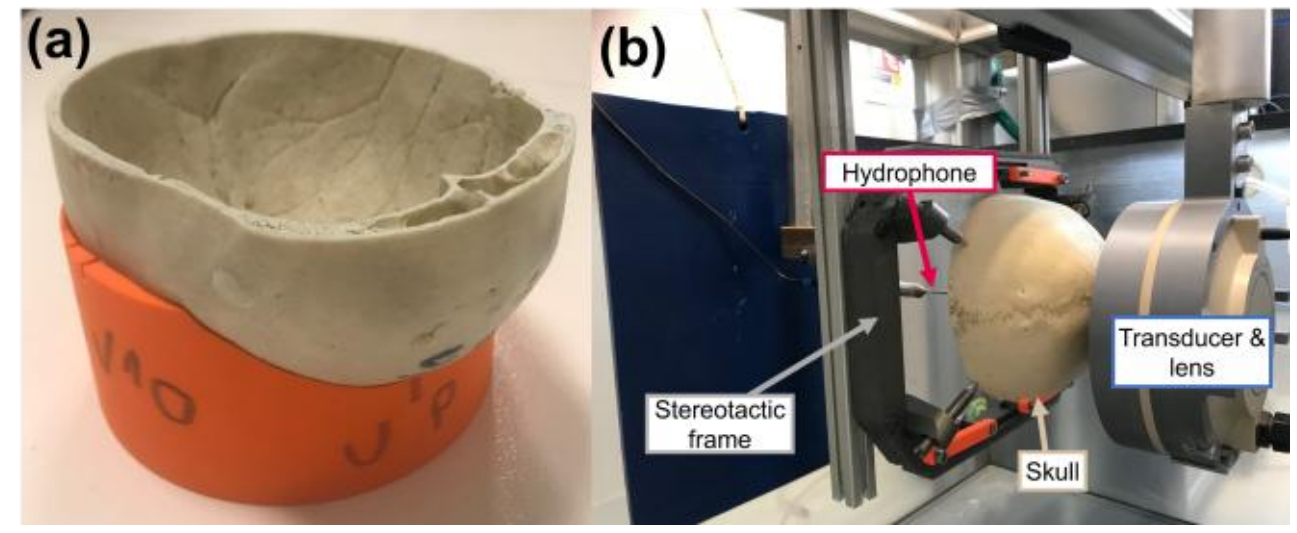

Fig. 1: Pictures of the experimental setup. (a) The skull placed in its custom 3D holder used to control its alignment with the transducer/lens assembly. (b) Full view of the experimental setup and display of the main items.

The lenses were designed to produce the two focal points at the focal depth of the transducer behind the skull. No axial steering was performed in addition to the lateral shift, so we evaluated the lens design with a 2D scan in the same plane. The dimensions of the acquisition planes in the direction of target separation were 27 mm and 42 mm for targets separated by 15 and 30 mm, respectively. In the other dimension, the plane was 12 mm large, regardless of the lateral distance between the targets (see Fig. 2). The scan resolution was 0.6 mm (one-fifth of the wavelength) in both directions. Each scan was then linearly interpolated to 0.3 mm (one-tenth of the wavelength) for a better visualization and higher precision on the computed metrics.

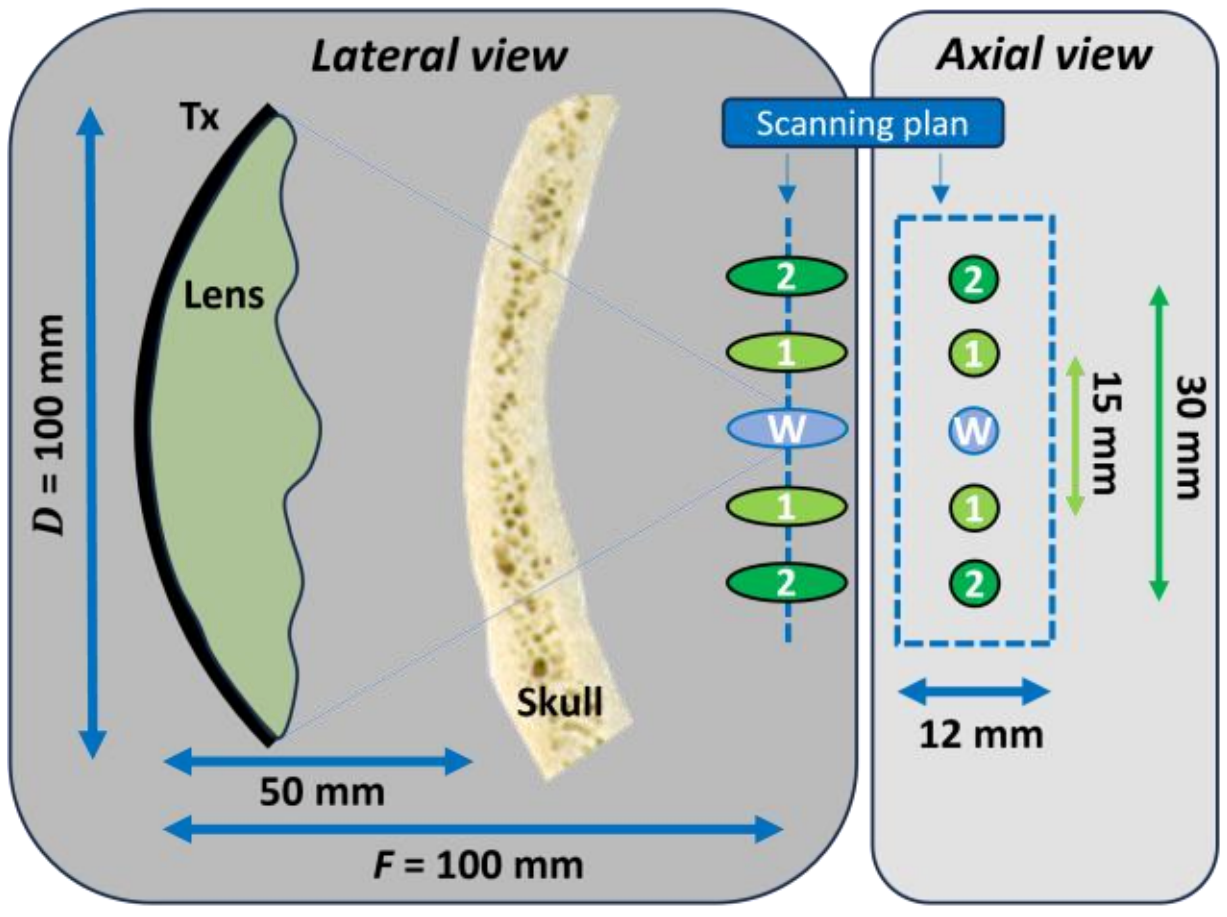

Fig. 2: Diagram of experimental setup. The position of the scan is shown in both lateral and axial views. Focal points with W represent the acoustic focal points in water. Focal points with 1 correspond to the case of 15 mm separation distance and focal points with 2 correspond to the case of 30 mm separation distance.

### E. Quality metrics definition

Several metrics were used to assess the quality of bilateral focusing. Spatial accuracy was assessed by measuring the difference (in mm) between the location of the maximum pressure obtained for each focal spot and the intended position of the targets. The maximum acoustic intensity in a central zone between the two targets (excluding a 1.5 wavelength radius around each target or 4.5 mm) was measured and expressed as a percentage of the maximum intensity measured over the entire 2D pressure field. This metric is related to the precision and safety of the treatment as it penalizes acoustic energy deposited outside the target areas (it will be referred as “safety metric”). Finally, another metric based on energy deposition was also calculated. The reference used to calculate this metric is the ideal bifocal case of time reversal in water (Fig. 5, grey area). The -6 dB (half pressure) surface of the time reversal pressure field corresponds to the engagement surface: it corresponds to the desired energy deposition pattern. We then calculated the percentage area of the planned engagement surface that overlapped with the -6 dB area of the lens-corrected pressure field. It quantifies the fraction of the half-intensity isodose that is actually delivered at the intended location. This metric is thus related to the efficacy of the treatment. It will be referred as “efficacy metric”.

## III. Results

Twenty acoustic lenses were experimentally evaluated in this study, through five different skulls, either with raw phase or with unwrapped phase, for each of the 15 mm and 30 mm lateral separation distance and lenses. The main finding is that the unwrapped phase lenses outperformed the non-unwrapped lenses in every metric calculated in this study. A lateral distance separation of 45 mm was also examined on one skull. This was used to illustrate the increasing number of phase wraps with increasing lateral distance and to discuss the achievable limits in terms of lateral distance separation (see Discussion section). Examples of 2D comparisons between raw phase and unwrapped phase lenses are shown for the 15 mm and 30 mm lateral distance between the targets in Fig. 3 and Fig. 4 respectively. Fig. 4.a highlights the appearance of a secondary lobe at the position of the acoustic focal point in water at almost half the maximum pressure measured at the targets. This unwanted lobe is removed when using the unwrapped phase lens, as can be seen in Fig. 4.(b) and (c). In some skulls, this central lobe may actually receive more energy than the target regions. This is highlighted in the axial field comparison on the main axis between the two targets across the five skulls for the two lateral target separations shown in Fig. 5.

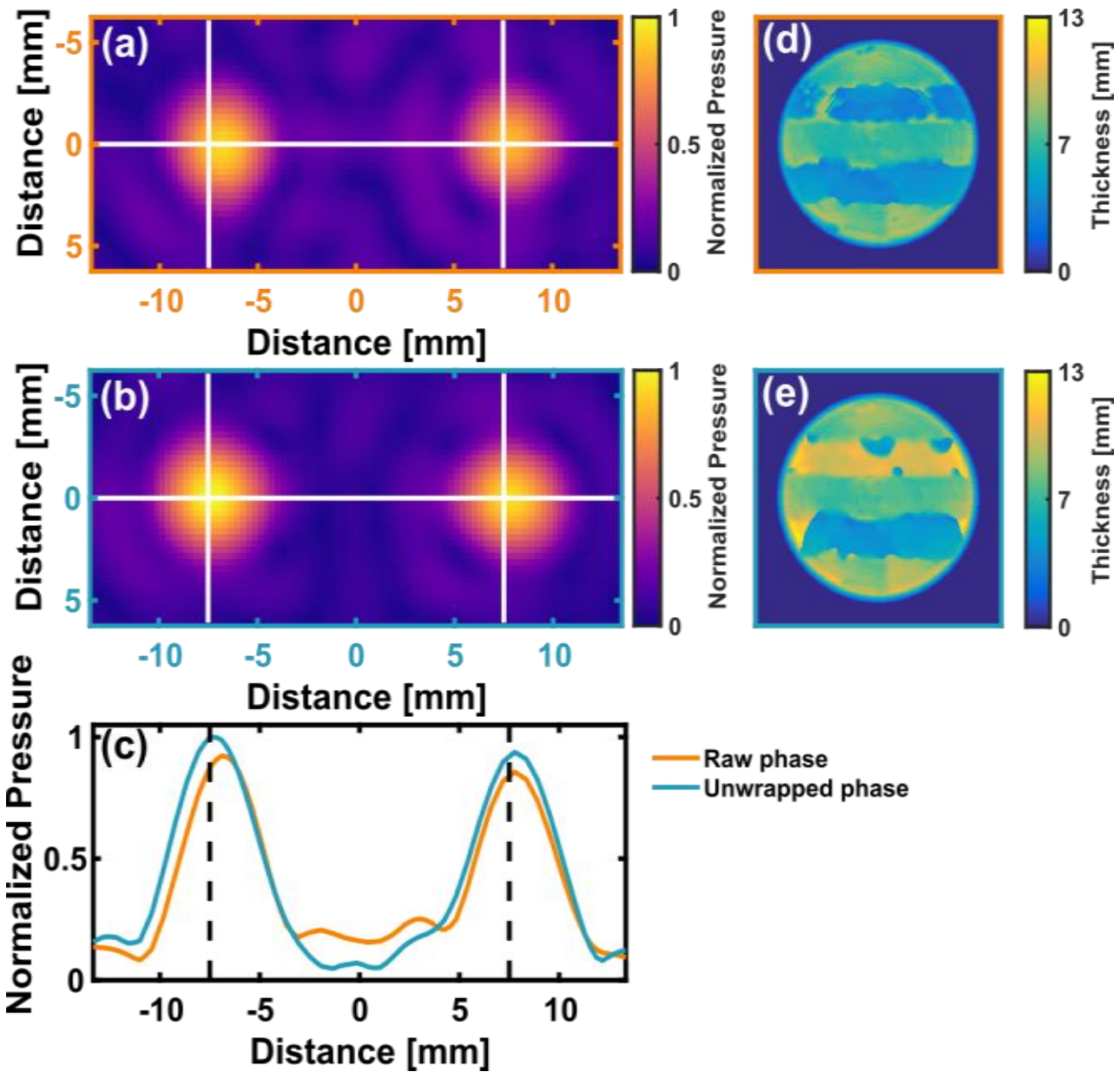

Fig. 3: (a)-(b) Experimental pressure fields obtained for bifocal lenses (15 mm between targets) with and without phase unwrapping (lens profile shown in (d) and (e) respectively) in skull #5. (c) Axial field comparison on the principal axis between the two targets. The pressures displayed are normalised to the maximum pressure value measured across (a) and (b).

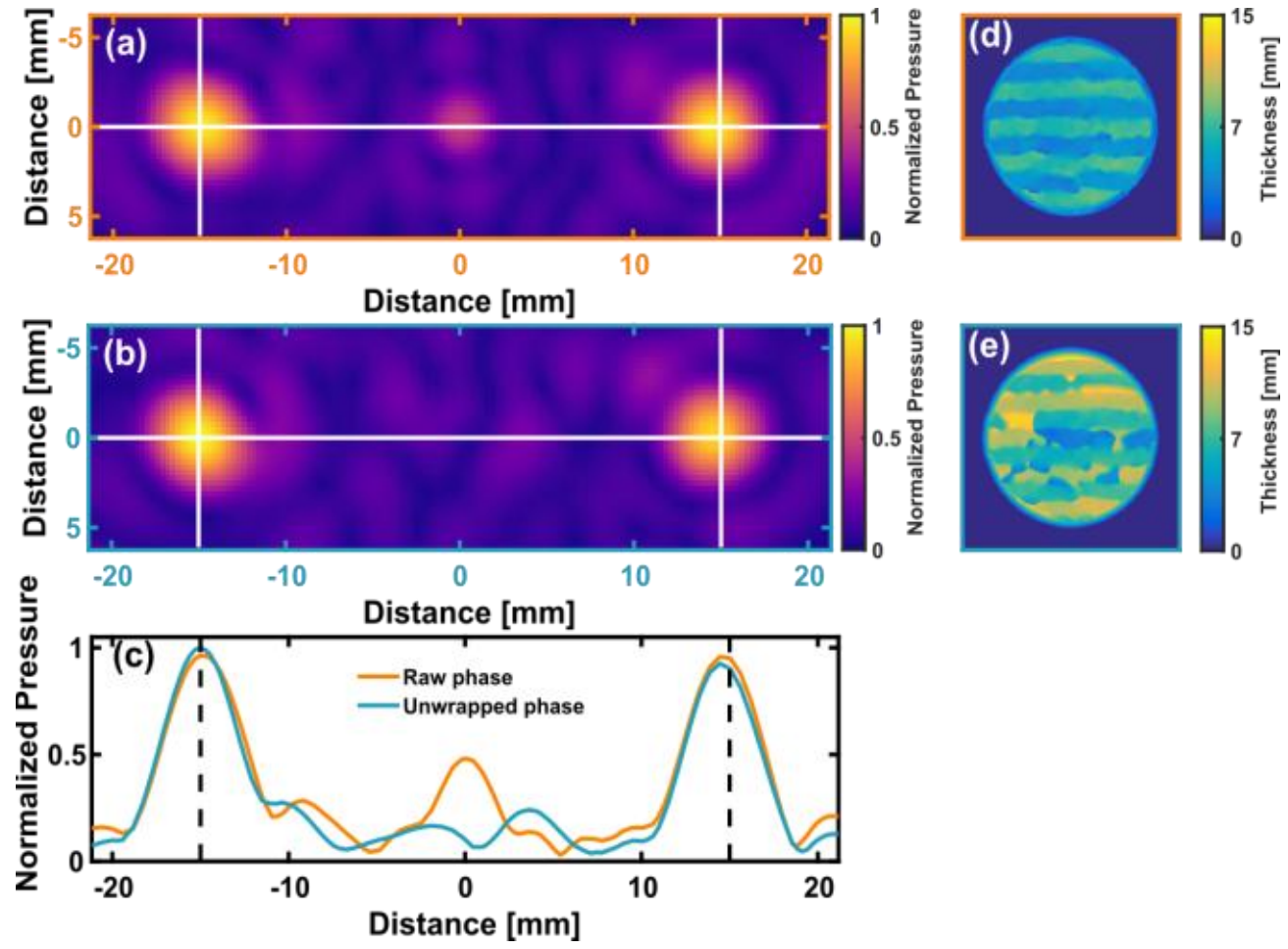

Fig. 4: (a)-(b) Experimental pressure fields obtained for bifocal lenses (30 mm between targets) with and without phase unwrapping (lens profile shown in (d) and (e) respectively) in skull #4. (c) Axial field comparison on the principal axis between the two targets. The pressures displayed are normalised to the maximum pressure value measured across (a) and (b).

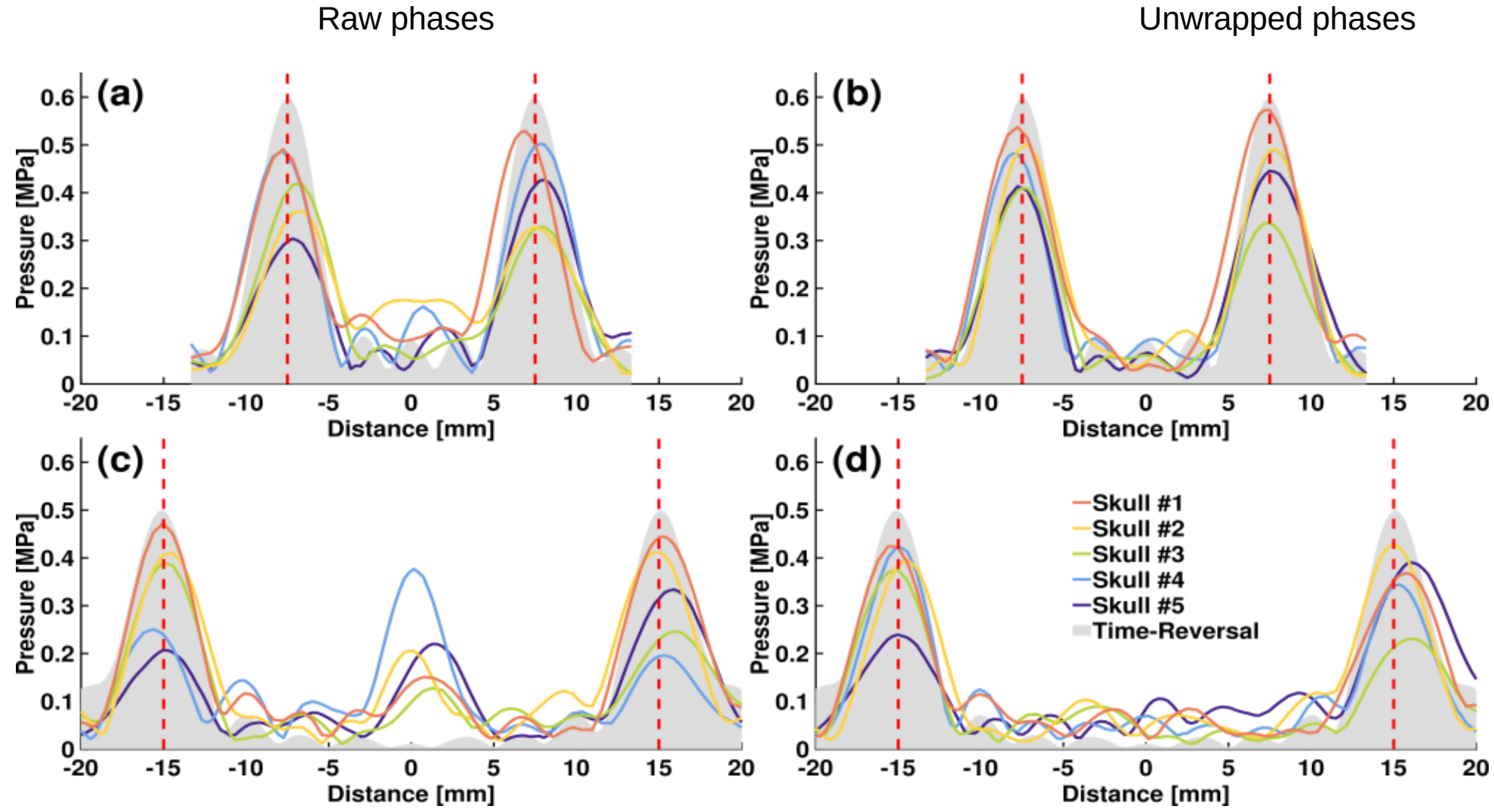

Fig. 5: Axial pressure field measurement on the principal axis between the two targets across the five skulls. (a)-(b) For a 15 mm lateral distance between targets (a) with raw phases and (b) when unwrapping the phase. (c)-(d) For a 30 mm lateral distance between targets (c) with raw phases and (d) when unwrapping the phase.

| | Distance between targets | Error in location of spatial peak pressure (mm) | Maximum intensity outside targets (%) "safety metric" | -6dB surface within the planed engagement surface (%) "efficacy metric" |
|---|---|---|---|---|
| *raw phase* | 15 mm | 0.4 ± 0.2 (WOP: 0.9) | 13.0 ± 8.9 (WOP: 28.61) | 89.7 ± 8.0 (WOP: 76.56) |
| *unwrapped phase* | 15 mm | 0.20 ± 0.2 (WOP: 0.6) | 5.1 ± 2.1 (WOP: 8.6) | 97.6 ± 1.6 (WOP: 95.0) |
| *Improvement with unwrapping (%)* | 15 mm | 35.8 ± 24.6 (BI: 55.6) | 51.4 ± 29.6 (BI: 82.2) | 9.4 ± 8.9 (BI: 24.1) |
| | | | | |
| *raw phase* | 30 mm | 3.2 ± 6.2 (WOP: 15.08) | 37.6 ± 37.4 (WOP: 100) | 73.9 ± 26.4 (WOP: 33.5) |
| *unwrapped phase* | 30 mm | 0.4 ± 0.4 (WOP: 1.2) | 8.3 ± 1.3 (WOP: 10.4) | 82.2 ± 13.9 (WOP: 62.8) |
| *Improvement with unwrapping (%)* | 30 mm | 73.5 ± 41.2 (BI: 96.2) | 58.5 ± 28.6 (BI: 91.2) | 28.5 ± 70.8 (BI: 155.1) |

Table I: Mean, standard deviation and worst observed performance of the calculated metrics on N=5 skulls for 15mm and 30mm lateral distances, both for raw phase and unwrapped phase lenses. Below, WOP means "Worst Observed Performance" and BI means "Best Improvement".

Table I summarizes the results and confirms the improvement of all the quality metrics when phase unwrapping is added to the design process. The average error in location between the intended and measured locations of the maximum pressure was reduced from 0.4 mm to 0.2 mm and from 3.3 mm to 0.4 mm using phase unwrapping for the 15 mm and 30 mm distances respectively. Most importantly, the worst performance observed, corresponding to Skull #2 and for the 30 mm lateral distance, was reduced from 14.8 mm to 0.3 mm. In this particular case, the location of the maximum pressure is not at any of the intended targets, but at the location of the center of curvature of the probe. The percentage of maximum intensity deposited outside the targets (safety metric) decreases from 13% to 5% for the 15 mm distance and from 38% to 8% for the 30 mm distance when the phase is unwrapped. The standard deviation of this metric is also drastically reduced when using phase unwrapping and drops from 9% to 2% and from 37% to 1% respectively, indicating an increased repeatability of the desired energy deposition. Finally, the surface engagement of the targets (efficacy metric) is on average increased by 9% and 28%.

In all cases, not only are the averages reduced, but the worst observed performance of each metric is also reduced when unwrapping is used, preventing undesirable stimulation patterns. For example, the worst observed performance for the efficacy metric with the raw phase and a 30 mm distance corresponds to skull #2. In this configuration, the 34% of the -6dB surface is within the planned engagement surface without phase unwrapping, compared to 86% with phase unwrapping, corresponding to a 155% increase in efficacy.

These metrics show that phase unwrapping reduces the likelihood of unwanted energy deposition in unwanted regions while improving target engagement. Unwrapped lenses achieve this with low variability and high reproducibility, a significant improvement in treatment safety and efficacy.

## IV. Discussion

As can be seen from Fig. 3.d, Fig. 4.d and Fig. 6.d, the number of phase wraps required to produce the bifocal pattern increases as the lateral distance between the targets increases. Our results also highlight a degradation in the performance of lenses with targets at increasing separation distance without phase unwrapping.

The absolute pressure on the targets is slightly increased by unwrapping the phase (50 kPa and 22 kPa increase on average for the 15mm and 30mm distances respectively, see Fig. 6). Thus, increased lens thickness after phase unwrapping does not significantly change the absolute pressure on the targets. This is consistent with the results obtained when comparing raw and unwrapped phase lenses for particle manipulation (see Supplementary Material of [46]). Acoustic absorption in a material is proportional to the thickness of material. The additional absorption due to the increased lens thickness after unwrapping appears to be balanced by the improved focusing capabilities. As the acoustic absorption increases linearly this frequency, this observation may not hold when the frequency used is increased or when a more attenuating material is used for the lens.

Increasing the lateral distance between the targets increases the number of phase jumps, and thus the number of abrupt changes in lens thickness: from 4 to 14 phase jumps with a distance between the targets from 15mm (Fig. 3.d) to 45mm respectively (Fig. 6. d.). In the case of the 45mm distance, an unwanted secondary lobe located at center of curvature of the transducer (distance = 0mm on Fig. 6 c.) as high as half the measured maximum pressure is successfully suppressed when the phase is unwrapped. A similar phenomenon can be observed with a 30mm distance (Fig. 4. c.).

A possible cause is a higher impact of refraction in unwrapped lenses. Indeed, when the number of phase wraps increases, the number of abrupt changes in lens thickness increases accordingly. Each abrupt change refracts and scatters the incoming acoustic wave. However, refraction and scattering are not taken into account in the design process because phase maps (unwrapped or not) are converted to local lens thickness by a formula that treats each lens pixel as an ideal phase shifter. Nevertheless, it remains to be demonstrated if diffraction induced by the abrupt changes are responsible for the appearance of a secondary lobe at the center of curvature of the transducer without unwrapping. First of all, phase wraps are still present in unwrapped lenses as the unwrapping is not perfect (Fig. 3.e; Fig. 4.e and Fig. 6.e). Second it can be noted that the number of horizontal abrupt thickness changes for the 15mm distance is the same (Fig. 3.d. and 3.e.). For the 45mm distance, for which the impact on the refocusing is more pronounced (Fig 6.c.), the overall shape of the lens is smoother but block patterns can be observed (Fig. 6e.).

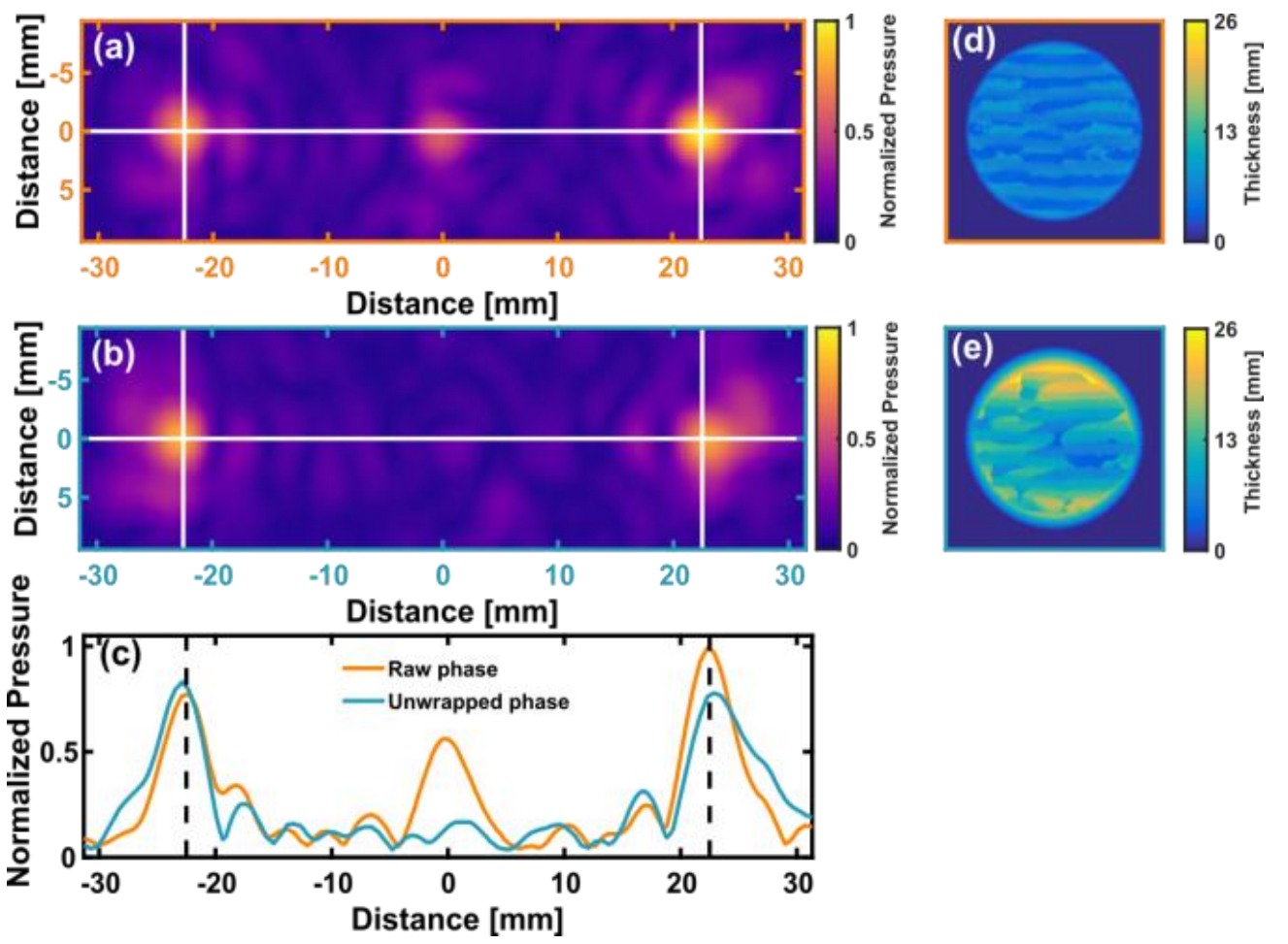

Fig. 6: (a)-(b) Experimental pressure fields obtained for bifocal lenses (45 mm between targets) with and without phase unwrapping (lens profile shown in (d) and (e) respectively) in skull #5. (c) Axial field comparison on the principal axis between the two targets. The pressures displayed are normalised to the maximum pressure value measured across (a) and (b).

The relative difference in pressure level between the two targets increases as the lateral distance between the targets increases, even though the phase is unwrapped (maximum difference is 20% for the 15 mm unwrapped case and up to 40% for the 30 mm unwrapped phase case (see Fig. 5. b Skull #3 and Fig. 5. d Skull #5 respectively). This could lead to uneven neural stimulation of the two brain targets, which could affect the outcome of the treatment and the interpretation of its success or failure. Iterative simulation-based optimization could improve the focusing for large lateral distances. This is beyond the current work and would require an extension of the work previously introduced in homogenous media [32] and would benefit from the reduction of the simulation time required for transcranial simulations [47].

The whole conclusion of this study can be extrapolated to the lateral steering of a single focal point. Indeed, increasing the lateral steering distance of a single focal point also increases the number of phase jumps and can benefit from phase-unwrapping, especially for broadband lens-based focusing [48]. On the other hand, binary acoustic lenses with abrupt thickness shifts have been used to dynamically change the position of the acoustic spot when the driving frequency is changed [22], [33] and it would be worth testing the impact of phase unwrapping in this particular case. Phase unwrapping reduces the number of phase and thickness discontinuities, allowing the associated refraction and scattering to be limited. It is worth noting that phase unwrapping can be used independently of the formula that is used to convert the phase shifts into the local thickness of the lens [21], [22].

## V. Conclusion

Overall, unwrapping the phase wraps computed during the design increases the fidelity of the bifocal pressure field generated across five human skulls. Previously introduced by Maimbourg et al. for monofocal lenses [19], [20], this study confirms that phase unwrapping is an efficient and low-computational cost method to enhance the focusing quality of acoustic lenses for transcranial applications. This work paves the way to a safe and reliable use of bifocal acoustic lenses for ultrasonic deep-brain stimulation in humans.

## Acknowledgments

The authors thank the brain imaging facility team (Prof. Catherine Oppenheim, Clément Debacker, Sylvain Charron, Maliesse Lui, Anna Simler Fayolle) of the GHU Paris Psychiatrie & Neurosciences for the CT images acquisition. We thank Image Guided Therapy (Pessac, France) and Imasonic (Voray-sur-l'Ognon, France) for their support.

This work was supported in part by the Agence Nationale de la Recherche under the program Ultrastim under Grant ANR-20-CE19-0013 and in part by the Focused Ultrasound Foundation Center of Excellence Program.